\documentclass[preprint]{aastex}
\usepackage{fullpage,verbatim,color}

\newcommand{\bdv}[1]{\mbox{\boldmath$#1$}}

\begin{document}

\title{MOA-2011-BLG-293Lb: A test of pure survey microlensing
planet detections}

\author{
J.C.~Yee\altaffilmark{1,2},
Y. Shvartzvald\altaffilmark{1,3},
A. Gal-Yam\altaffilmark{1,4},
I.A. Bond\altaffilmark{5,6},
A. Udalski\altaffilmark{7,8},
S. Koz{\l}owski\altaffilmark{1,7,8},
C. Han\altaffilmark{1,9},
A. Gould\altaffilmark{1,2},
J. Skowron\altaffilmark{1,7,2},
D.~Suzuki\altaffilmark{5,10},\\
and\\
F.~Abe\altaffilmark{11}, 
D.P.~Bennett\altaffilmark{12},
C.S.~Botzler\altaffilmark{13}, 
P. Chote\altaffilmark{14},
M.~Freeman\altaffilmark{13}, 
A.~Fukui\altaffilmark{15}, 
K.~Furusawa\altaffilmark{11}, 
Y.~Itow\altaffilmark{11},  
S. Kobara\altaffilmark{11},
C.H.~Ling\altaffilmark{6},
K.~Masuda\altaffilmark{11}, 
Y.~Matsubara\altaffilmark{11},
N.~Miyake\altaffilmark{11}, 
Y. Muraki\altaffilmark{11},
K. Ohmori\altaffilmark{11},
K.~Ohnishi\altaffilmark{16},
N.J.~Rattenbury\altaffilmark{17},
To.~Saito\altaffilmark{18},
D.J.~Sullivan\altaffilmark{14}, 
T.~Sumi\altaffilmark{10}, 
K. Suzuki\altaffilmark{11}, 
W.L.~Sweatman\altaffilmark{6}, 
S. Takino\altaffilmark{11}, 
P.J.~Tristram\altaffilmark{19}, 
K. Wada\altaffilmark{10} \\
(the MOA Collaboration),\\
M.\,K. Szyma{\'n}ski\altaffilmark{8}, 
M. Kubiak\altaffilmark{8}, 
G. Pietrzy{\'n}ski\altaffilmark{8,20}, 
I. Soszy{\'n}ski\altaffilmark{8}, 
R. Poleski\altaffilmark{8}, 
K. Ulaczyk\altaffilmark{8}, 
{\L}. Wyrzykowski\altaffilmark{8,21}, 
P. Pietrukowicz\altaffilmark{8}\\
(the OGLE Collaboration),\\
W. Allen\altaffilmark{22},
L. A. Almeida\altaffilmark{23},
V. Batista\altaffilmark{2},
M. Bos\altaffilmark{24},
G. Christie\altaffilmark{25},
D.L. DePoy\altaffilmark{26},
Subo Dong\altaffilmark{27}\footnote{Sagan Fellow},
J. Drummond\altaffilmark{28},
I. Finkelman\altaffilmark{3},
B.S. Gaudi\altaffilmark{2},
E. Gorbikov\altaffilmark{3},
C. Henderson\altaffilmark{2},
D. Higgins\altaffilmark{29},
F. Jablonski\altaffilmark{23},
S. Kaspi\altaffilmark{3},
I. Manulis\altaffilmark{4,30},
D. Maoz\altaffilmark{3},
J. McCormick\altaffilmark{31},
D. McGregor\altaffilmark{2},
L.A.G Monard\altaffilmark{32},
D. Moorhouse\altaffilmark{33},
J.A. Mu\~{n}oz\altaffilmark{34},
T. Natusch\altaffilmark{25,35},
H. Ngan\altaffilmark{25},
E. Ofek\altaffilmark{4},
R.W. Pogge\altaffilmark{2},
R. Santallo\altaffilmark{36},
T.-G. Tan\altaffilmark{37},
G. Thornley\altaffilmark{33},\\
and\\
I.-G. Shin\altaffilmark{9},
J.-Y. Choi\altaffilmark{9},
S.-Y. Park\altaffilmark{9},
C.-U. Lee\altaffilmark{38},
J.-R. Koo\altaffilmark{38}\\
(the $\mu$FUN Collaboration)
}
\altaffiltext{1}{Microlensing Follow Up Network ($\mu$FUN)}
\altaffiltext{2}{Department of Astronomy, Ohio State University, 140
W. 18th Ave., Columbus, OH 43210, USA;
gaudi, gould, henderson, jyee, pogge@astronomy.ohio-state.edu}
\altaffiltext{3}{School of Physics and Astronomy and Wise
Observatory, Tel-Aviv University, Tel-Aviv 69978, Israel;
shai, dani, shporer, david@wise.tau.ac.il}
\altaffiltext{4}{Department of Particle Physics and Astrophysics,
Weizmann Institute of Science, 76100 Rehovot, Israel}
\altaffiltext{5}{Microlensing Observations in Astrophysics (MOA) Collaboration}
\altaffiltext{6} {Institute of Information and Mathematical
Sciences, Massey University, Private Bag 102-904, North Shore Mail
Centre, Auckland, New Zealand;
i.a.bond, c.h.ling, w.sweatman@massey.ac.nz}
\altaffiltext{7}{Optical Gravitational Lens Experiment (OGLE)}
\altaffiltext{8}{Warsaw University Observatory, Al. Ujazdowskie 4,
00-478 Warszawa, Poland; e-mail: udalski, msz, mk, pietrzyn, simkoz,
soszynsk, kulaczyk, rpoleski@astrouw.edu.pl}
\altaffiltext{9}{Department of Physics, Institute for Astrophysics,
Chungbuk National University, Cheongju 371-763, Korea}

\altaffiltext{10}{Department of Earth and Space Science, Osaka University, Osaka 560-0043, Japan.}
\altaffiltext{11}{Solar-Terrestrial Environment Laboratory, Nagoya University, Nagoya, 464-8601, Japan}
\altaffiltext{12}{University of Notre Dame, Department of Physics, 225 Nieuwland Science Hall, Notre Dame, IN 46556-5670 USA;bennett@nd.edu} 
\altaffiltext{13}{Department of Physics, University of Auckland, Private Bag 92-019, Auckland 1001, New Zealand}
\altaffiltext{14}{School of Chemical and Physical Sciences, Victoria University, Wellington, New Zealand}
\altaffiltext{15}{Okayama Astrophysical Observatory, National Astronomical Observatory
of Japan, Asakuchi, Okayama 719-0232, Japan}
\altaffiltext{16}{Nagano National College of Technology, Nagano 381-8550, Japan}\altaffiltext{17}{Jodrell Bank Observatory, The University of Manchester, Macclesfield, Cheshire SK11 9DL, UK}
\altaffiltext{18}{Tokyo Metropolitan College of Aeronautics, Tokyo 116-8523, Japan}
\altaffiltext{19}{Mt. John University Observatory, P.O. Box 56, Lake Tekapo 8770, New Zealand}

\altaffiltext{20}{Universidad de Concepci\'{o}n, Departamento de Astronomia,
    Casilla 160--C, Concepci\'{o}n, Chile}
\altaffiltext{21}{Institute of Astronomy, University of Cambridge,
    Madingley Road, Cambridge CB3 0HA, UK e-mail:
    wyrzykow@ast.cam.ac.uk }
\altaffiltext{22}
{Vintage Lane Observatory, Blenheim, New Zealand; whallen@xtra.co.nz}
\altaffiltext{23}{Instituto Nacional de Pesquisas Espaciais, S\~{a}o
Jos\'{e} dos Campos, SP, Brazil}
\altaffiltext{24}
{Molehill Astronomical Observatory, North Shore, New Zealand}
\altaffiltext{25}{Auckland Observatory, Auckland, New Zealand; gwchristie@christie.org.nz}
\altaffiltext{26}
{Dept.\ of Physics, Texas A\&M University, College Station, TX, USA; 
depoy@physics.tamu.edu}
\altaffiltext{27}{Institute for Advanced Study, Einstein Drive,
Princeton, NJ 08540, USA; dong@ias.edu}
\altaffiltext{28}{Possum Observatory; john\_drummond@xtra.co.nz}
\altaffiltext{29}{Hunters Hill Observatory, Canberra, Australia}

\altaffiltext{30}{The Davidson Institute of Science Education,
Weizmann Institute of Science, 76100 Rehovot, Israel}
\altaffiltext{31}
{Farm Cove Observatory, Centre for Backyard Astrophysics,
Pakuranga, Auckland, New Zealand; farmcoveobs@xtra.co.nz}
\altaffiltext{32}
{Klein Karoo Observatory, Centre for Backyard Astrophysics, Calitzdorp, South
Africa; bmonard@mweb.co.za}
\altaffiltext{33}{Kumeu Observatory, Kumeu, New Zealand; acrux@orcon.net.nz, guy.thornley@gmail.com}
\altaffiltext{34}{Departamento de Astronom\'{i}a y Astrof\'{i}sica, Universidad de Valencia, E-46100 Burjassot, Valencia, Spain}
\altaffiltext{35}{AUT University, Auckland, New Zealand; tim.natusch@aut.ac.nz}
\altaffiltext{36}
{Southern Stars Observatory, Faaa, Tahiti, French Polynesia; santallo@southernstars-observatory.org}
\altaffiltext{37}{Perth Exoplanet Survey Telescope, Perth, Australia}
\altaffiltext{38}{Korea Astronomy and Space Science Institute,
Daejeon 305-348, Korea}

\begin{abstract}
Because of the development of large-format, wide-field cameras,
microlensing surveys are now able to monitor millions of stars with
sufficient cadence to detect planets. These new discoveries will span
the full range of significance levels including planetary signals too
small to be distinguished from the noise. At present, we do not
understand where the threshold is for detecting
planets. MOA-2011-BLG-293Lb is the first planet to be published from
the new surveys, and it also has substantial followup
observations. This planet is robustly detected in survey+followup data
($\Delta\chi^2\sim5400$). The planet/host mass ratio is $q=5.3\pm
0.2\times 10^{-3}$. The best fit projected separation is $s=0.548\pm
0.005$ Einstein radii. However, due to the $s\leftrightarrow s^{-1}$
degeneracy, projected separations of $s^{-1}$ are only marginally
disfavored at $\Delta\chi^2=3$. A Bayesian estimate of the host mass
gives $M_L = 0.43^{+0.27}_{-0.17}\,M_\odot$, with a sharp upper limit
of $M_L < 1.2\,M_\odot$ from upper limits on the lens flux.  Hence,
the planet mass is $m_p=2.4^{+1.5}_{-0.9}M_{\rm
Jup}$, and the physical projected separation is either $r_\perp \simeq
1.0\,$AU or $r_\perp \simeq 3.4\,$AU. We show that survey data alone
predict this solution and are able to characterize the planet, but the
$\Delta\chi^2$ is much smaller ($\Delta\chi^2\sim500$) than with the
followup data. The $\Delta\chi^2$ for the survey data alone is smaller
than for any other securely detected planet. This event suggests a
means to probe the detection threshold, by analyzing a large sample of
events like MOA-2011-BLG-293, which have both followup data and high
cadence survey data, to provide a guide for the interpretation of
pure survey microlensing data.

\end{abstract}

\section{Introduction}

Large-format, wide-field cameras have placed microlensing on the cusp
of joining RV and transits as a technique able to find dozens of
planets at a time \citep{Shvartzvald12}, moving the field from the
discovery of individual objects to the study of planet
populations. Using these new cameras, ``second generation''
microlensing surveys will be able to effectively monitor an order of
magnitude more events for anomalies due to planets. At the same time,
they can maintain an observing strategy that makes no reference to
whether or not a planetary signal is suspected, thus enabling a
statistically robust sample of events whose detection efficiencies are
well understood. One requirement for such a sample is that all events
must be analyzed for planets, including signals at the limits of
detectability. At present the detection threshold is poorly understood
since the current practice is only to analyze the most obvious
signals. However, it is known that microlensing data have systematics
and correlated noise that make it difficult to use standard
statistical measures to set the detection thresholds. In this paper,
we analyze the microlensing event MOA-2011-BLG-293, which is covered
by all three second-generation survey telescopes and also has
substantial followup data, and we suggest a means to study the
boundary of what is detectable in the second generation surveys.

Originally, the purpose of microlensing surveys was simply to identify
ongoing microlensing events, which requires monitoring several million
stars with a cadence of about once per night. Because a typical
planetary signal lasts for only a few hours, it was nearly impossible
to detect planets from the early survey data. Thus, in order to detect
planets, higher cadence followup data were needed\footnote{Implicit in
this is that the data quality is good enough for planet detection.}. One
followup strategy is to monitor one or more targets with increased
cadence to provide additional coverage of the light curve and to
search for anomalies. A second strategy is continuous or
near-continuous monitoring of a single event of interest, usually
because it is suspected to be high magnification ($A_{\rm max}> 100$)
or anomalous. These additional observations can be taken either by
dedicated followup groups or the surveys themselves can go into
followup mode (typically, continuous or near-continuous observations)
if they deem an event to be of sufficient interest. Therefore, in
followup mode, both survey and followup groups may modify their
target list and/or observing cadence in response to suspected
planetary signals.  This strategy has been effective at finding
planets but makes understanding the detection efficiencies complex,
although this has been done successfully for high magnification events
in \citet{Gould10}. Additionally, \citet{Sumi10} were able to derive a
slope (but not the normalization) for the mass ratio function of
planets from the planetary events known at the time. Of the 13
microlensing planets published to date\footnote{For completeness, we
note that there are at least two other planets claimed in the
literature, which are not considered secure detections. A possible
circumbinary planet proposed to explain anomalies in MACHO 97-BLG-41
\citep{Bennett99} can also be explained by orbital motion of the
binary lens \citep{Albrow00}. There is also evidence for a planetary
companion to the lens in MACHO 98-BLG-35, but only with
$\Delta\chi^2=20$ (\citealt{Gaudi02} contains a discussion of why this
is inadequate for detection).}, only one was published from data taken
in a pure survey mode \citep{Bennett08}\footnote{At least one other
pure survey detection should be published soon
\citep[see][]{Bennett12}.}.

The new high-cadence, systematic surveys will have sufficient cadence
and data quality to detect and characterize planets with masses as
small as the Earth without additional followup data
\citep{Gaudi08b}. Such pure-survey detections require near-24-hour
monitoring with a cadence of several observations per hour. Many of
these future discoveries will be part of a rigorous experiment wherein
the detection efficiencies are well understood because they will be
found in blind or blinded (in which followup data are removed)
searches. High cadence surveys, even without global coverage, also
allow additional science such as the detection of very short timescale
events \citep{Sumi11}. Recent upgrades by the Optical Gravitational
Lensing Experiment (OGLE; Chile) and Microlensing Observations in
Astrophysics (MOA; New Zealand) collaborations augmented by the Wise
Observatory (Israel) survey now allow near-continuous monitoring
(observations every 15-30 minutes) of several fields in the Galactic Bulge,
22 hours/day \citep{Shvartzvald12}.

The power of these new surveys comes from the combination of
high-cadence, systematic observations, which were previously only
achievable through followup for a small subset of events, and the
ability to monitor millions of stars. At the same time, followup
observations maintain some advantages over current surveys. Because
the followup networks have access to additional telescopes at various
sites, followup observations often have redundancy.  This makes them
less vulnerable to bad weather, which can create gaps in the
data. Additionally, multiple data sets at a given epoch provide a
check on systematics or other astrophysical phenomena that may create
false microlensing-like signals \citep[see][]{Gould12}. Simultaneous
or near simultaneous observations from multiple sites are also
required to measure terrestrial microlens parallax
\citep[e.g.][]{Gould09}. Furthermore, since followup observations are
targeted, they can achieve a much higher cadence, and are frequently
continuous, although the current strategy for surveys is typically to
switch to near continuous followup observations for events of
interest. Finally, while followup groups routinely make an intensive
effort to get observations in additional filters\footnote{Microlensing
observations are normally done in $I$-band (or similar filters)
because that is the optical band that is most sensitive toward the
Galactic Bulge. In order to derive source colors as in
Sec. \ref{sec:cmd}, we need observations in a different filter, for
which we typically use $V$-band.}, survey groups are less aggressive
about obtaining such observations. All of these additional bits of
information can increase confidence in the microlensing interpretation
and reduce ambiguity in the models. The trade off is that with
somewhat sparser data coverage, surveys are able to systematically
monitor more than an order of magnitude more events.

The additional planets detected by surveys, which are not currently
being detected with followup, will fall into two categories related to
the two kinds of caustics produced by a 2-body lens. First, there will
be many more planetary caustic anomalies detected. These caustics are
created along or near the planet-star axis at a distance from the lens star
that depends on the projected separation, $s$, and the mass ratio,
$q$, between the two bodies. Anomalies created by these caustics can
be found with current followup but since the source trajectory is
random with respect to the binary axis, these anomalies occur in a
random place in the light curve. Hence, surveys will detect more of
them because they can observe more stars. Second, a planetary
companion to the lens induces a caustic at the position of the lens
star, the so-called ``central'' caustic. Source crossings of such
caustics can be predicted in advance because they require that the
source trajectory pass very close to the position of the lens star.
Surveys will observe more central caustic events that are too faint to
observe with current followup or are not recognized to be high
magnification quickly enough to organize followup observations. For
both types of events, there is the question of whether the survey data
alone are indeed sufficient to detect planets in individual
microlensing light curves in spite of having sparser data. For central
caustic events there is an additional question of whether or not the
anomaly will be sufficiently characterized, since the models can be
quite degenerate for these kinds of events, sometimes with little
constraint on the mass ratio between the lens star and its companion
\citep[e.g.][]{Choi12}. Given this much larger sample of events which
will contain signals of all significance levels including ones that
can be confused with systematics, the challenge is to create a subset
of events for which the vast majority of planetary signals can be
considered reliable, and secondarily for which the planets are
well-characterized, in the case of central caustic-type
events. \citet{Gould10} estimate a threshold of
$\Delta\chi^2=$350--700 would be appropriate, but the true value is
unknown. In principle, such questions could be addressed with
simulations. However, in simulations it is difficult to account for
real effects such as data systematics and stellar variability. Hence,
using actual microlens data provides field testing that complements
results from simulations.

MOA-2011-BLG-293 provides an opportunity for investigating survey-only
detection thresholds. The planet is robustly detected in the
survey+followup data ($\Delta\chi^2\sim5400$), and the event was
observed by all three current survey telescopes. Wise Observatory
obtained data of the anomaly in their normal survey mode without
changing their observing cadence, and the rest of the light curve is
reasonably well covered by OGLE and MOA survey data. For this event,
we are able to determine whether the survey data alone can
successfully ``predict'' the solution determined when all of the data
are included.

This event also has the faintest source of any published planetary
event. We show that for such faint sources small systematic errors in
the flux measurements can radically affect the microlensing solution,
even when all the anomalous features occur at high magnification when
the source is bright. In particular, the source flux and the event
timescale are determined primarily from data near baseline where small
systematic errors may be of order the change in flux being measured.
Because the systematic errors in the timescale propagate to many other
quantities including the planet/star mass ratio, they must be
investigated carefully. This is particularly important for future
surveys where many of the events will be at or beyond the magnitude
limit at baseline.

We begin by presenting the discovery and observations of
MOA-2011-BLG-293 in Section \ref{sec:data}. The color-magnitude
diagram of the event is presented in Section \ref{sec:cmd} and used to
derive the intrinsic source flux. In Section \ref{sec:ulenslc}, we
address the consequences of systematics in the measured flux when they
are similar in magnitude to the source flux. Then, in Section
\ref{sec:analysis} we present the analysis of the light curve of the
event, and we compare the results with and without followup data in
Section \ref{sec:surveyonly}. Additional properties of the event are
derived in Section \ref{sec:eventprop}, and the physical properties of
the lens star and planet are derived from a Galactic model in Section
\ref{sec:physical}. We discuss the implications for future survey-only
detections in Section \ref{sec:discuss}. Finally, the possibility of
detecting the lens with adaptive optics (AO) observations is discussed
in the Appendix.

{\section{Data Collection and Reduction}
\label{sec:data}}

MOA issued an electronic alert for MOA-2011-BLG-293 [(RA,Dec) =
(17:55:39.35, $-$28:28:36.65), (l,b)=(1.52,$-1.66$)] at UT 10:27, 4
Jul 2011 (HJD$'=\,$HJD-2450000 = 5746.94), based on survey
observations from their 1.8m telescope with a broad $R/I$ filter and
2.2 deg$^2$ imager at Mt.\ John, New Zealand.  At UT 12:45, the
Microlensing Follow-Up Network ($\mu$FUN) refit the data and announced
that this was a possible high-magnification event, where
``high-magnification'' is $A_{\rm max}\gtrsim100$. At UT 17:28,
$\mu$FUN upgraded to a full high-magnification alert ($A_{\rm
max}>270)$, emailing subscribers to their email alert service, which
includes members of $\mu$FUN and other microlensing groups, to urge
observations from Africa, South America, and Israel. Additionally, a
shortened version of the alert was posted to Twitter. This prompted
$\mu$FUN Weizmann to initiate the first followup observations at UT
19:45, using their 0.4m telescope ($I$ band) at the Martin S. Kraar
Observatory located on top of the accelerator tower at the Weizmann
Institute of Science Campus in Rehovot, Israel. At UT 23:25, $\mu$FUN
Chile initiated continuous observations using the SMARTS 1.3m
telescope at CTIO.  At UT 00:00 $\mu$FUN issued an anomaly alert based
on the first four photometry points from CTIO, which were rapidly
declining when the expected behavior was rapid brightening. The great
majority of the CTIO observations were in $I$ band, but seven
observations were taken in $V$ band to measure the source color.  In
addition, the SMARTS ANDICAM camera takes $H$ band images
simultaneously with each $V$ and $I$ image.  These are not used in the
light curve analysis but are important in the Appendix.

MOA-2011-BLG-293 lies within the survey footprint of the MOA, OGLE,
and Wise microlensing surveys and so was scheduled for ``automatic
observations'' at high cadence at all three observatories.  MOA
observed this event at least 5 times per hour. Wise observed this
field 10 times during the 4.6 hours that it was visible from their
1.0m telescope, equipped with 1 deg$^2$ imager and $I$-band filter, at
Mitzpe Ramon, Israel. The event lies in OGLE field 504, one of three
very high cadence fields, which OGLE would normally observe about 3
times per hour.  In fact, it was observed at a much higher rate, but
with the same exposure time, in response to the high-magnification alert and
anomaly alert.  Unfortunately, high winds prevented opening of the
telescope until UT 01:02. OGLE employs the 1.3m Warsaw telescope at
Las Campanas Observatory in Chile, equipped with a 1.4 deg$^2$ imager
primarily using an $I$-band filter.

\begin{figure}
\includegraphics[width=\textwidth]{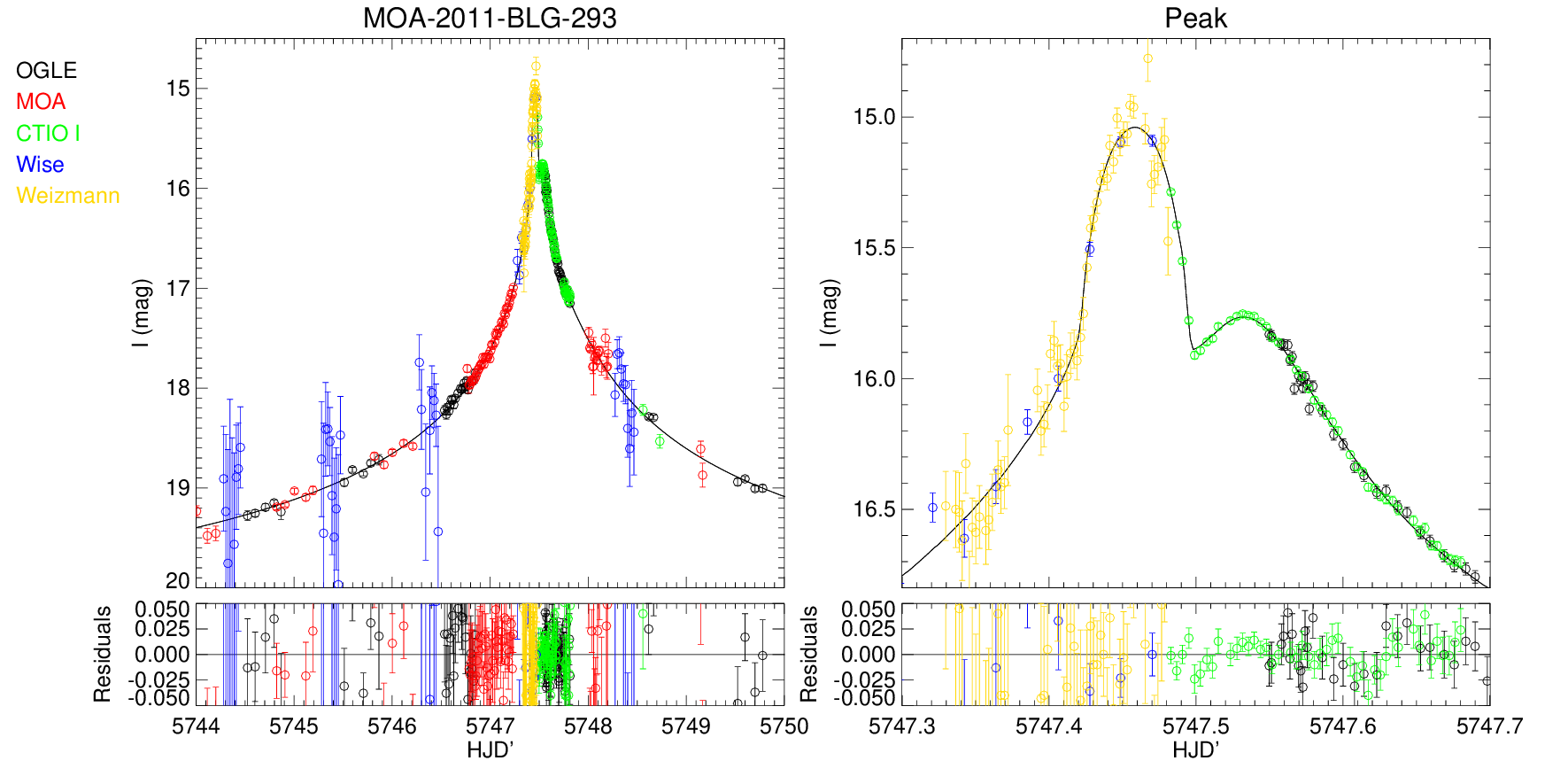}
\caption{The light curve of MOA-2011-BLG-293. The left-hand panel
shows a broad view of the light curve, while the right-hand panel
highlights the peak of the event where the planetary perturbation
occurs. Data from different observatories are represented by different
colors, see legend. The black curve is the best-fit model with a close
topology ($s<1$). The times are given in
HJD$^{\prime}$=HJD$-2450000$. \label{fig:lc}}
\end{figure}

\begin{figure}
\includegraphics{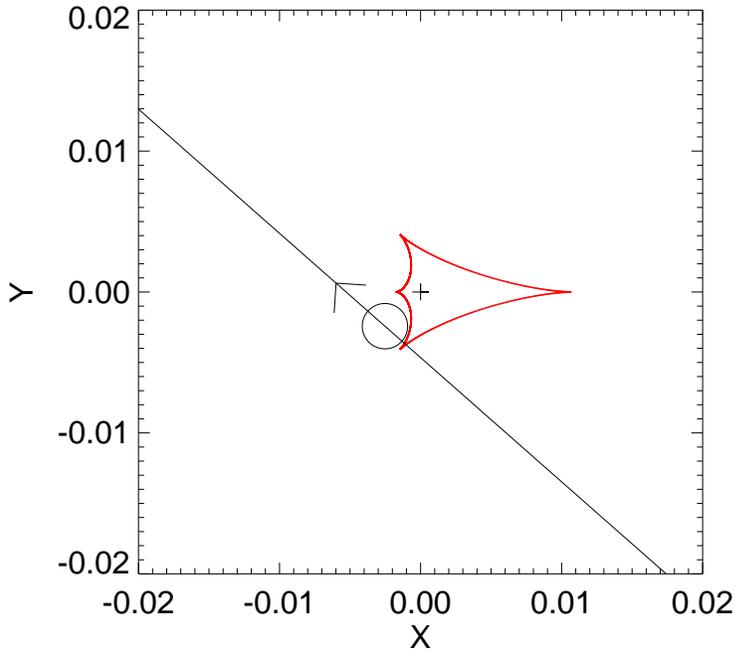}
\caption{Caustic structure and source trajectory of the best-fit model
of MOA-2011-BLG-293 in the source plane. The circle shows the physical
size of the source, and its position at the time of the caustic exit
($HJD^{\prime}\sim5747.5$). The x-axis is the star-planet axis, and
the origin is at the center of magnification. The scale of the axes is
in units of the Einstein radius.\label{fig:traj}}
\end{figure}

The data are shown in Figure \ref{fig:lc}.  Several features should be
noted.  First, the pronounced part of the anomaly lasts just 4 hours
beginning at HJD$'=5747.40$.  The main feature is quite striking,
becoming about one magnitude brighter in about one hour.  The coverage
during the anomaly is temporally disjoint between the observatories in
Israel and those in Chile, a point to which we return below.  Finally,
the CTIO data show a discontinuous change of slope (``break''), which
is the hallmark of a caustic exit, when the source passes from being
partially or fully inside a caustic to being fully outside the
caustic (see Fig. \ref{fig:traj}).

MOA and OGLE data were reduced using their standard pipelines
\citep{Bond01,Udalski03} which are based on difference image analysis
(DIA). In the case of the OGLE data, the source is undetected in the
template image. Since the OGLE pipeline reports photometry in
magnitudes, an artificial blend star with a flux of 800 units
($I_{OGLE}=20.44$) was added to the position of the event to prevent
measurements of negative flux (and undefined magnitudes) at baseline
when the source is unmagnified.

Data from the remaining three observatories were also reduced using
DIA \citep{Wozniak00}, with each reduction specifically adapted to
that imager. Using comparison stars, the Wise and Weizmann photometry
were aligned to the same flux scale as the CTIO $I$ band
by inverting the technique of \citet{Gould10a}.  That is, the
instrumental source color was determined from CTIO observations, and
then the instrumental flux ratios (CTIO vs.\ Wise, or CTIO vs.\
Weizmann) were measured for field stars of similar color.  The
uncertainties in these flux alignments are 0.016 mag for Wise and 0.061 mag
for Weizmann.

{\subsection{Data Binning and Error Normalization}
\label{sec:error}}

Since photometry packages typically underestimate the true errors,
which have a contribution from systematics, we renormalize the error
bars on the data, as is done for most microlensing events. After
finding an initial model, we calculate the cumulative $\chi^2$
distribution for each set of data sorted by magnification. We
renormalize the error bars using the formula
\begin{equation}
\label{eqn:errors}
\sigma_i^{\prime} = k\sqrt{\sigma_i^2+e_{\rm min}^2}
\end{equation}
and choosing values of $k$ and $e_{\rm min}$ such that the $\chi^2$
per degree of freedom $\chi^2_{\rm red}=1$ and the cumulative sum of
$\chi^2$ is approximately linear as a function of source
magnification. Specifically, we sort the data points by magnification,
calculate the $\Delta\chi^2$ contributed by each point, and plot
$\sum^N_i \Delta\chi^2_i$ as a function of $N$ to create the cumulative
sum of $\chi^2$, where $N$ is the number of points with magnification
less than or equal to the magnification of point $N$. Note that
$\sigma_i$ is the uncertainty in magnitudes (rather than flux). The
values of $k$ and $e_{\rm min}$ for each data set are given in Table
\ref{tab:data}. Except for OGLE, the values of $e_{\rm min}$ are all
zero. This term compensates for unrealistically small uncertainties in
the measured magnitude, which can happen when the event is bright and
the Poisson flux errors are small.

For the MOA data, we eliminate all observations with $t$ outside the interval
$5743.5<t(HJD^{\prime})<5749.5$ (see Section \ref{sec:tefs}). We also
exclude all MOA points with seeing $>5^{\prime\prime}$ because these
data show a strong nonlinear trend with seeing at baseline. After
making these cuts, we renormalize the data as described above.

To speed computation, the OGLE and MOA data in the wings of the event
were binned.  In the process of the binning, $3\sigma$ outliers were
removed. This binning does not account for correlations in the data,
which if they exist can increase the reduced $\chi^2$ above the
nominal value of $\chi^2_{\rm red}=1$.

\section{CMD}
\label{sec:cmd}
\begin{figure}
\includegraphics[width=\textwidth]{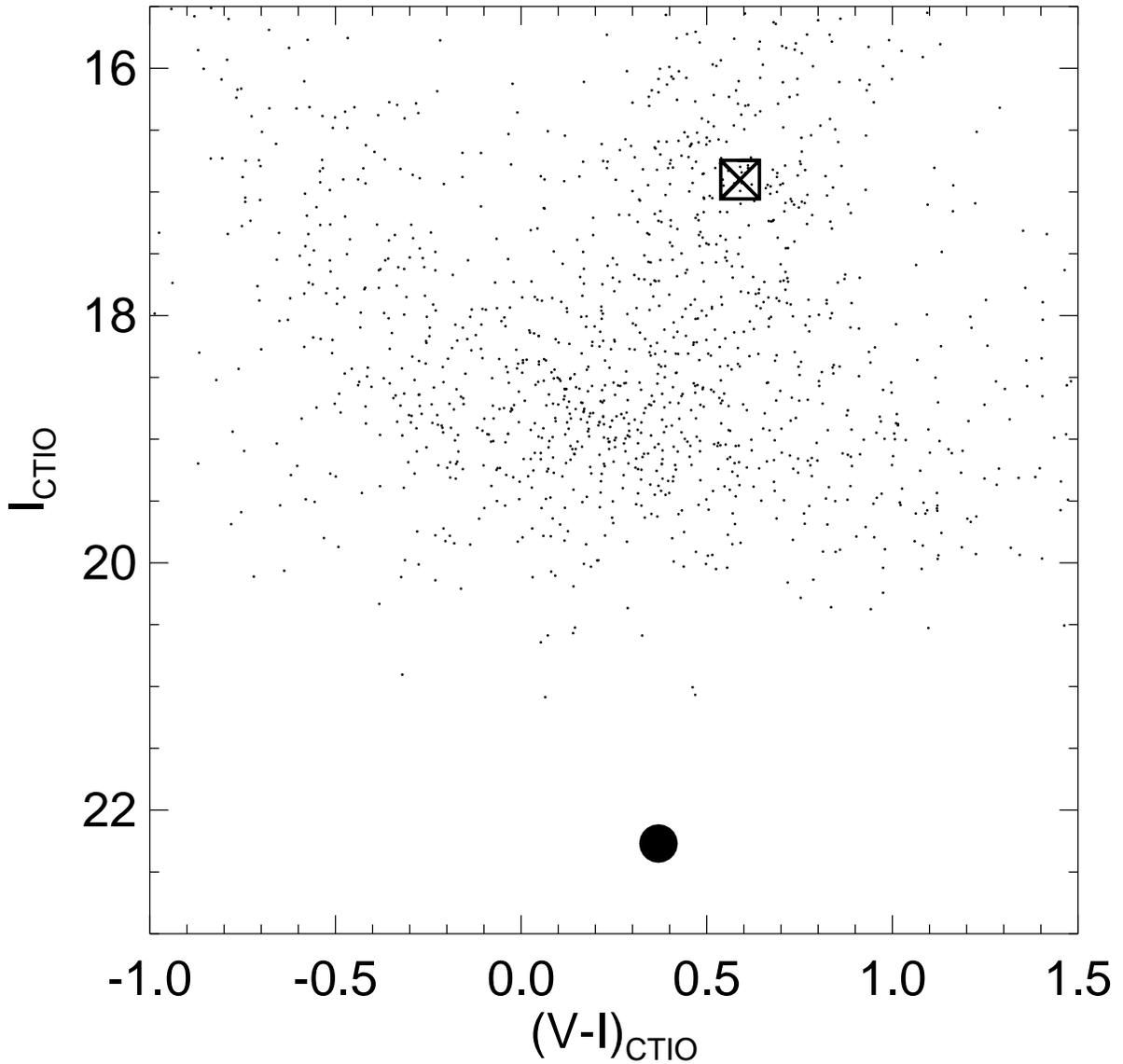}
\caption{Color-Magnitude Diagram of the event in instrumental
(uncalibrated) magnitudes. The source is shown as the solid black
point; the errors in the source color and magnitude are smaller than
the size of the point. The centroid of the Red Clump is the open
square with an X through it. The small points show the stars in
the field, restricted to stars within 60$^{\prime\prime}$ of the
source because there is strong differential reddening on larger
scales. \label{fig:cmd}}
\end{figure}

We use the CTIO $I$ and $V$ band data to
construct a color-magnitude diagram (CMD) of the event. We measure the
instrumental (uncalibrated) source color by linear regression of the
$V$ and $I$ fluxes (which is independent of the model) and the
magnitude from the $f_{S,{\rm CTIO}}$ of our best-fit model:
$(V-I,I)_S=(0.37,22.27)\pm(0.03,0.05)$. The position of the source
relative to the field stars within 60$''$ of the source (small dots)
is shown in Figure \ref{fig:cmd} as the solid black dot. We calibrate
these magnitudes and account for the reddening toward the field by
assuming the source is in (or at least suffers the same extinction as)
the Bulge and using the centroid of the red clump as a standard
candle. Because of strong differential extinction across the field, we
use only stars within 60$''$ of the source to measure the centroid of
the red clump. Since the event is in a low latitude field, there are
more stars than is typical for bulge fields and the red-clump centroid
can be reliably determined even with this restriction. In instrumental
magnitudes, the centroid of the red clump is
$(V-I,I)_{cl}=(0.59,16.90)$ compared to its intrinsic value of
$(V-I,I)_{cl,0}=(1.06,14.32)$ \citep{Bensby11,Nataf12}, which assumes
a Galactocentric distance $R_0=8\,$kpc and that the mean clump
distance toward $l=1.5$ lies 0.1 mag closer than $R_0$
\citep{Rattenbury07}. We can apply the offset between these two values
to the source color and magnitude to obtain the calibrated, dereddened
values $(V-I,I)_{S,0}= (0.84,19.69)\pm(0.05,0.16)$. The uncertainty in
the color is derived from \citet{Bensby11} by comparing the
spectroscopic colors to the microlens colors of that sample. The
uncertainty in the calibrated magnitude is the sum in quadrature of
the uncertainty in $f_{S,{\rm CTIO}}$ from the models (0.05 mag), the
uncertainty in $R_0$ (5\%$\rightarrow$0.1 mag), the uncertainty in the
intrinsic clump magnitude (0.05 mag), and the uncertainty in
centroiding the red clump (0.1 mag).

{\section{Effect of Faint Sources on Microlens Parameters}
\label{sec:ulenslc}}

The source star in MOA-2011-BLG-293 is extremely faint with an
apparent magnitude in the OGLE photometry of $I_{S,{\rm
OGLE}}=21.7$. Consequently, the measured flux errors can be comparable
to or larger than the source flux, particularly near baseline. Because
of this, systematics in the baseline data must be carefully
accounted for so as not to bias the microlens results. Systematics in
the measured flux at the level of $f_S$ can lead to biases in the
measured Einstein timescale, $t_{\rm E}$, of the same order. We begin by
discussing robust parameters, which can be measured solely from
the highly magnified portion of the light curve and so, are
independent of uncertainties in the flux measured near baseline. Then in
Section \ref{sec:tefs}, we discuss in detail the effect of systematics
in the measured baseline flux on the microlens parameters,
particularly $t_E$ and the mass ratio between the components of the
lens, $q$.

{\subsection{Robustly Measured Parameters}
\label{sec:invariants}}

At high magnification, a microlensing light curve for a point source
being lensed by a point lens can be described by the unmagnified,
baseline flux of the event, $f_{\rm base}$, and three parameters
(``invariants''): the time of the peak, $t_0$, the difference between
$f_{\rm base}$ and the peak flux, $f_{\rm lim}$, and the effective
width of the light curve, $t_{\rm eff}$ \citep[see Eq. (2.4) and (2.5)
in]{Gould96}. These parameters are nearly invariant under changes to
the source flux and are robustly determined by the light curve. The
change in the observed flux due to the event can then be written as a
function of these invariants:
\begin{equation}
f_{\rm obs}(t) - f_{\rm base} = G_3(t; t_0, t_{\rm eff}, f_{\rm lim}),
\end{equation}
where $f_{\rm obs}$ is the observed flux and the subscript on $G$
refers to the number of parameters. Note that $f_{\rm base}$ is also
an observable.

In the limit where the event is highly magnified, the exact form of
$G_3(t)$ can be derived from the microlens variables.  The three
microlens variables of a point-source--point-lens microlensing model
are $t_0$, the impact parameter in units of the Einstein radius, $u_0$, and the
Einstein crossing time, $t_{\rm E}$. The observed flux is given by
\begin{equation}
\label{eqn:flux}
f_{\rm obs}=f_SA(t)+f_B=f_S[A(t)-1]+f_{\rm base},
\end{equation}
where $A(t)$ is the magnification of the source, $f_S$ is the flux of
the source, and $f_B$ is the flux of all other stars blended into the
PSF (including the flux from the lens). By definition, $f_{\rm base} =
f_S+f_B$. For a point lens in the limit of high magnification
($A(t)\gg1$), the magnification is given by
\begin{equation}
\label{eqn:mag}
A(t)\simeq\frac{1}{u_0Q(t)}
\end{equation}
where
\begin{equation}
Q(t; t_0, t_{\rm eff}) = \sqrt{1+\left(\frac{t-t_0}{t_{\rm eff}}\right)^2},
\end{equation}
is a function of only time and the invariant:
\begin{eqnarray}
t_{\rm eff}\equiv u_0t_{\rm E}.
\end{eqnarray}
In this limit, the evolution of the observed flux, $G_3(t;t_0,t_{\rm
eff},f_{\rm lim})$, is then given by
\begin{equation}
\label{eqn:fHM}
G_3(t; t_0, t_{\rm eff}, f_{\rm lim})= \frac{f_{\rm lim}}{Q(t)}, \mathrm{\,where\,} f_{\rm lim}\equiv \frac{f_S}{u_0}.
\end{equation}

If finite source effects are detected, the change in the observed flux
is a more complicated function because of the additional microlens
parameter $\rho$, which is the source size in units of the Einstein
radius. However, there is also an additional invariant
\begin{equation}
t_{\star}\equiv\rho t_E,
\end{equation}
the source crossing time, which determines the width of the peak of
the light curve for a point lens. Hence, the change in the observed
flux can be written as
\begin{equation}
f_{\rm obs}(t) - f_{\rm base} = G_4(t; t_0, t_{\rm eff}, f_{\rm lim},t_{\star})
=G_3(t)B\left(Qt_{\rm eff}/t_{\star}\right),
\end{equation}
where $B(Qt_{\rm eff}/t_{\star})=B(u/\rho)$ is a function composed of
elliptic integrals, whose exact form is derived in \citet{Gould94} and
\citet{Yoo04b}.

In the case of a two-body lens like MOA-2011-BLG-293, the invariants
may not be obvious from the light curve, but they are still robustly
measured as we show below. For example, the width of the peak is
distorted by two-body perturbation, but based on the source trajectory
(Fig. \ref{fig:traj}), it can be seen that the width of the first bump
at HJD$^{\prime}5747.46$, which is caused by the cusp crossing, will
be slightly larger than $2t_{\star}$. For a two-body lens with a
central caustic crossing, there is also another invariant due to the
mass ratio, $q$, between the two lensing bodies, so
\begin{equation}
f_{\rm obs}(t) - f_{\rm base} = G_5(t; t_0, t_{\rm eff}, f_{\rm lim},
t_{\star}, qt_E).
\end{equation}
This new invariant $q t_{\rm E}$ can be understood as follows. For
central caustics\footnote{Two-body lenses with unequal mass ratios
will create one caustic at the position of the more massive body, the
``central'' caustic, and another set of caustics elsewhere, the
``planetary'' caustics.}, like the one in MOA-2011-BLG-293, the
caustic size is proportional to the mass ratio of the two lensing
bodies, $q$, and the caustic shape is roughly constant for a given
$s$. Therefore, the time between successive features in the light
curve is set by $q t_{\rm E}$, i.e., the size of the caustic
multiplied by the characteristic timescale, and since the observed
times of the features can be well measured, the uncertainty in this
quantity is extremely small. In this case, the main features are the
two bumps in the light curve and the discontinuity in the slope that
occurs between the bumps. [Note that a two-body lens introduces two
parameters in addition to $q$: the separation between the two bodies
projected onto the plane of the sky, $s$, and the angle of the source
trajectory with respect to the binary axis, $\alpha$. Because these
parameters are of less interest, we do not discuss them in this
context.]

Table \ref{tab:invariants} shows that for this event these quantities,
$t_{\rm eff}$, $f_{\rm lim}$, $t_{\star}$, and $q t_{\rm E}$, do
indeed have extremely small uncertainties and can approximately be
considered invariants.

{\subsection{Parameters Vulnerable to Systematics}
\label{sec:tefs}}

The above invariants are determined by the data taken near the peak of
the light curve where $A\gg1$. However, in order to extract the values
of the microlens parameters, $u_0$, $\rho$, and $q$, from the
invariants, we must measure $t_{\rm E}$. The information on $t_E$ must
necessarily come from the wings of the light curve where the
magnification is small and $A(t)\neq1/u(t)$ \citep{Dominik09}. Since
the magnification is small, the change in the observed flux compared
to the source flux is also small, so the measurement of $t_E$ may be
considered to be the statistical sum of many measurements of a small
change in flux. In order to get an accurate measurement of $t_{\rm
E}$, the statistical {\it and} systematic errors in the flux must be
smaller than the change we are trying to measure.

\begin{figure}
\includegraphics{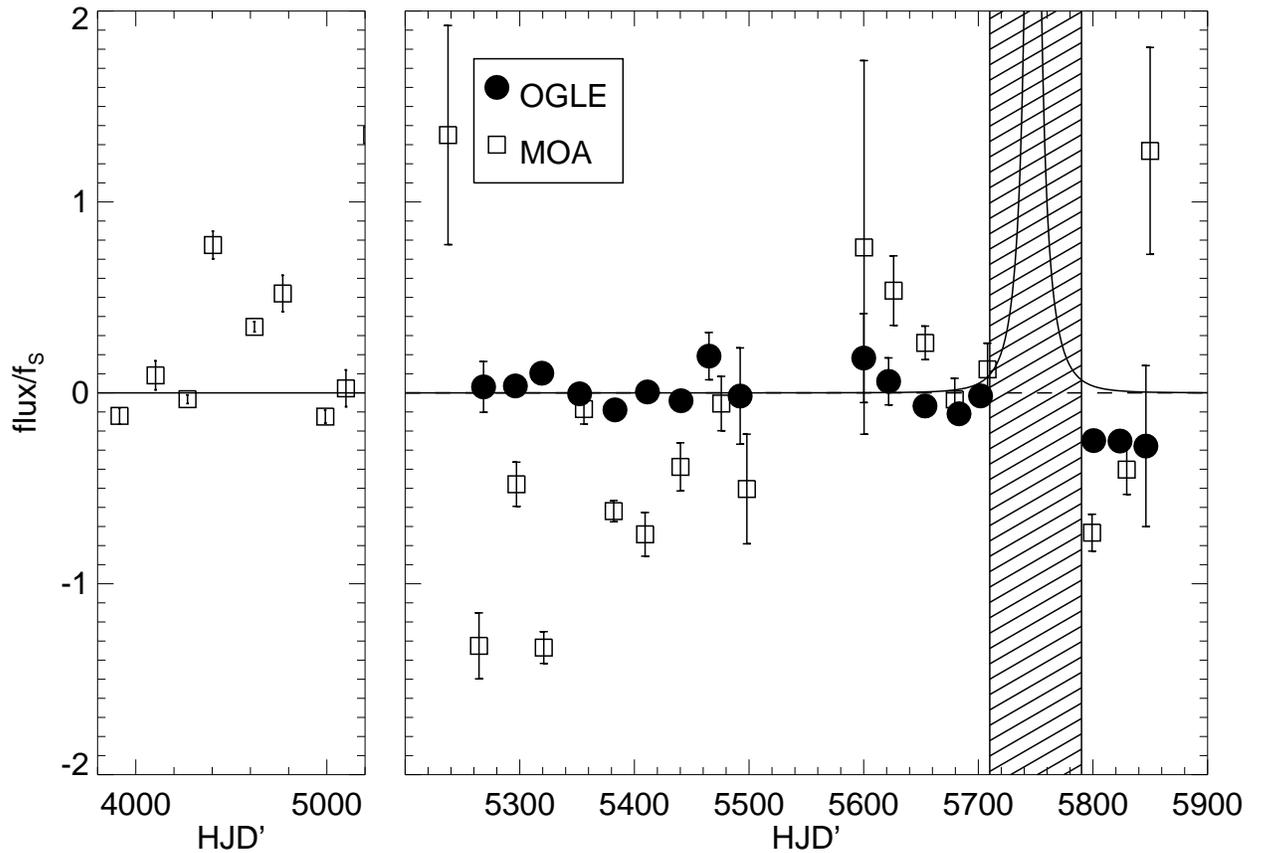}
\caption{Observed MOA (open squares) and OGLE (solid circles) fluxes
at baseline. The fluxes have been scaled by the source flux and
adjusted so that the baseline is approximately zero. The solid line
shows the expected flux from the model. The data have been binned by
30 days ({\it right panel}) and semi-annually ({\it left panel}). Data taken when the
event is significantly magnified (hashed region: $5710<t
(HJD^{\prime})<5790$) have been excluded. Note that the MOA data show
significant variation at a level comparable to the source
flux.\label{fig:fbase}}
\end{figure}

In the case of MOA-2011-BLG-293, because the source flux is extremely
faint, it is difficult to measure accurately. When the magnification
is a factor of a few or less (in the wings and at baseline), if the
flux is not measured with an accuracy substantially smaller than the
source flux, this can lead to bias in the measurement of $f_S$ or
equivalently $t_{\rm E}$, since $f_St_{\rm E}=f_{\rm lim}t_{\rm eff}$
is robustly determined, and so to bias in quantities dependent on $t_{\rm
E}$ such as $q$. To check for this possible source of bias, we bin the
OGLE and MOA data by 30 days to see if their measurements of the
baseline flux are stable (Fig. \ref{fig:fbase}). We find that the OGLE
measurement of the baseline flux is stable at a level that is smaller
than the observed source flux. Therefore, we use all of the OGLE data
in our models. We note that the flux after the event ($t > {\rm
HJD}'5790$) appears to be at a lower level than the baseline before
the event. In Section \ref{sec:fbase}, we discuss the effect of
assuming the baseline decreases at a constant rate during the course
of the event.

As seen in Figure \ref{fig:fbase}, the MOA baseline flux exhibits
scatter in excess of the measured photometric errors, and there is
also variation in measured baseline flux from season to season. The
magnitude of this scatter is similar to the magnitude of the source
flux.  Because of this variation, we conclude that the baseline flux
is not sufficiently well measured in the MOA data to detect the small
changes in flux necessary to measure $t_E$. As a result, to avoid
biasing our results, we use only the MOA data from the peak of the
light curve where the photometry is precise: $5743.5<t({\rm
HJD}')<5749.5$.

\section{Analysis}
\label{sec:analysis}

Without any modeling, we can make some basic inferences about the
relevant microlens parameters from inspection of the light
curve. MOA-2011-BLG-293 increases in brightness from $I\sim 19.7$ to
$I\sim 15.0$, indicating a source magnification of at least
75. Additionally, except for the deviations at the peak, the event is
symmetric about $t_0$. From these two properties, we infer that only
central or resonant caustics (both of which are centered on the position of the
primary) are relevant to the search for microlens models.

We fit the light curve using a Markov Chain Monte Carlo (MCMC)
procedure. In addition to the parameters described in Section
\ref{sec:ulenslc}, a model with a two-body lens has two additional
parameters: the angle of the source trajectory with respect to the
binary axis defined to be positive in the clockwise
direction\footnote{The binary axis has its origin at the center of
magnification and is positive in the direction of the planet.},
$\alpha$, and the projected separation between the two components of
the lens scaled to the Einstein radius, $s$. Because they are
approximately constants, we use the parameters $t_{\rm eff}$ and
$t_{\star}$ in place of the microlens variables $u_0$ and $\rho$. For
a given model, Equation (\ref{eqn:flux}) must be evaluated for each
observatory, $i$, so $f_S,f_B\rightarrow f_{S,i}, f_{B,i}$. We adopt
the ``natural'' linear limb-darkening coefficients $\Gamma= 2u/(3-u)$
\citep{Albrow99}.  Based on the measured position of the source in the
CMD, we estimate that $T_{\rm eff}=5315$K and $\log g = 4.5$ cgs. We
average the linear limb-darkening coefficients for $T_{\rm eff}=5250$K
and $T_{\rm eff}=5500$K from \citet{Claret00} assuming $v_{\rm
turb}=2\,$km s$^{-1}$ to find $\Gamma_V=0.6368$ and $\Gamma_I=0.4602$.

The magnifications are calculated on an $(s,q)$ grid, using the
``map-making'' technique \citep{Dong06} in the strong finite-source
regime and the ``hexadecapole'' approximation \citep{Pejcha09,Gould08}
in the intermediate regime.

We began by searching a grid of $s$ and $q$ to obtain a basic solution
for the light curve. For central caustic crossing events like this
one, there is a well known degeneracy between models with close
topologies ($s<1$) and wide topologies ($s>1$)
\citep[e.g.][]{Griest98}. We initially searched a broad grid for close
topologies and then used the results to inform our search for wide
solutions, since to first order, $s\rightarrow s^{-1}$. The basic
model from this broad grid has $s\sim0.55$, $q\sim0.005$, and
$\alpha\sim220^{\circ}$, such that the source passes over a cusp at
the back end of a central caustic. This caustic is created by a
two-body lens with a mass ratio similar to that of a massive Jovian
planet orbiting a star. Figure \ref{fig:traj} shows this basic
geometry with the source trajectory relative to the caustic
structure. The bump in the light curve at HJD$^{\prime}\sim 5747.45$
is created when the source passes over the cusp of the caustic.

Because the Wise and Weizmann data only overlap with other data sets
where their errors are extremely large, there is some concern that the
parameters of the models will be poorly constrained, since within the
standard modeling approach the flux levels of these data can be
arbitrarily adjusted up or down relative to the other data. However,
from the flux alignment described in Section \ref{sec:data}, we have
an estimate of $f_{S,i}$ for these data relative to $f_{S,{\rm
CTIO}}$. This alignment gives us an independent means to test the
validity of our model. If the model is correct, then the values of
$f_{S,{\rm Wise}}$ and $f_{S,{\rm Weizmann}}$ should agree with
$f_{S,{\rm CTIO}}$ within the allowed uncertainties. Alternatively, if
we include the flux-alignment constraint in the MCMC fits, the
solution should not change significantly.

We incorporate the flux-alignment constraint in a way that is
parallel to the model constraints from the data, i.e., by introducing a
$\chi^2$ penalty:
\begin{equation}
\chi^2_b =\sum_i\frac{(f_{S,{\rm CTIO}} - f_{S,i})^2}{\sigma_{{\rm flux},i}^2};
\quad
\sigma_{{\rm flux},i} = \frac{\ln 10}{2.5}
\left(\frac{f_{S,{\rm CTIO}}+f_{S,i}}{2}\right)\sigma_i,
\label{eqn:sigflux}
\end{equation}
where $i$ corresponds to the observatory with the constraint, and
$\sigma_i$ is the uncertainty in magnitudes of the flux alignment for
that observatory. In the absence of any constraints, the flux
parameters for each observatory, $f_{S,i}$ and $f_{B,i}$, are linear
and their values for a particular model can be found by inverting a
block-diagonal covariance matrix, $b$.  We include the flux
constraints by adding half of the second derivatives of $\chi^2_b$
to the $b$ matrix:
\begin{equation}
\Delta b(f_{S,i},f_{S,k}) = {2\delta_{ik} - 1 \over \sigma_{{\rm flux},i}^2},
\end{equation}
where $k=$CTIO and $\delta_{ik}$ is a Kronecker-delta.  This couples
formerly independent $2\times2$ blocks. Strictly speaking, the
equation for $\sigma_{{\rm flux},i}$ given in Equation (\ref{eqn:sigflux}) is
a numerical approximation. Therefore, we iterate the linear fit until
the value of $\sigma_{{\rm flux},i}$ is converged, which typically occurs in
only a few iterations.

We refined the $(s,q)$ grid around our initial close solution, fitting
the data both with and without flux-alignment constraints. The mean and
$1\,\sigma$ confidence intervals for the parameters from these two
fits are given in Table~\ref{tab:models}. There are only small
quantitative differences between the two solutions, and nothing that
changes the qualitative behavior of the model. The slight increase
in $\chi^2$ is expected because of the additional term due to the flux
constraints. After finding this close solution, we repeated the grid
with $s\rightarrow s^{-1}$ to identify the wide solution. The parameters
of this solution are also given in Table~\ref{tab:models} both with
and without flux-alignment constraints.  The close solution is mildly preferred
over the wide solution by $\Delta\chi^2\sim3$, so we quote the values
for the flux-constrained close solution:
\begin{equation}
q=5.3\pm 0.2\times 10^{-3}
\qquad
s = 0.548 \pm 0.005, 
\label{eqn:sols}
\end{equation}
noting that the two topologies give very similar solutions (except
$s\rightarrow s^{-1}$).

Additionally, we searched for a parallax signal in the event by adding
two additional free parameters to the fit for the close solution:
$\pi_{{\rm E},N}$ and $\pi_{{\rm E},E}$, the North and East components
of the parallax vector (e.g., \citealt{Gould04}). The parameters of
this fit are given in Table \ref{tab:models}. No parallax signal was
detected, and we found no interesting constraints on these
parameters. The $\chi^2$ improves for fits including parallax by only
$\Delta\chi^2=7$ for two additional degrees of freedom. In some cases,
even when parallax is not detected, meaningful upper limits can be
placed on the parallax, but in this case we have an uninteresting
3$\sigma$ constraint of $0\leq|\pi_{\rm E}|\leq 7.8$.

{\subsection{Effect of Systematics in $f_{\rm base}$}
\label{sec:fbase}}

As discussed in Section \ref{sec:tefs}, it is possible that the
underlying OGLE baseline flux is changing during the course of the
event. In order to test how that could introduce systematic effects in
our results, we create a fake OGLE data set accounting for a constant
decrease in baseline flux during the event. Specifically, we assume
that the baseline flux decreases at a constant rate between
HJD$^{\prime}5710.$ and HJD$^{\prime}5790.$ leading to an overall
decrease in flux of $0.27f_{S,{\rm OGLE}}$. We then repeat the MCMC procedure
for the close solution including flux-alignment constraints. We find that the
value of $t_E$ increases by 15\%, and consequently, the values of $q$,
$u_0$, and $\rho$ decrease by the same amount. In principle, this
could represent a systematic error in our results. However, at the
present time, the evidence for a change in the baseline flux is weak,
so we only report these results for the sake of completeness.

{\subsection{Analysis with Survey-Only Data}
\label{sec:surveyonly}}

From this analysis, we have a robustly detected planet
($\Delta\chi^2\sim 5400$ compared to a point lens\footnote{Note that
the numbers quoted for the point lens models include constraints from
the flux alignment in the fit. Removing the flux-alignment constraints
improves the $\chi^2$, primarily because the Weizmann data can be
scaled arbitrarily. However, compared to the planet fit, the point
lens fit without flux-alignment constraints is still extremely poor,
$\Delta\chi^2\sim4400$. Flux-alignment constraints have very little
effect on the point lens fit to survey only data.}) and a well-defined
solution. Now we can ask whether the planet could have been detected
from the survey data alone, whether the solution is well-constrained,
and most importantly, whether it is the same solution. To begin, we
construct a ``survey only'' subset of the data. We first eliminate the
Weizmann and CTIO data.  Second, we ``thin out'' the OGLE data to
mimic OGLE survey data as they would have been if there had been no
high-magnification or anomaly alerts.  OGLE data on several nights
previous to (and following) the peak have a cadence of 1 observation
per 0.015 days.  We therefore adopt a subset of 18 (out of 44) OGLE
points from the peak night with this sampling rate.

\begin{figure}
\includegraphics[width=\textwidth]{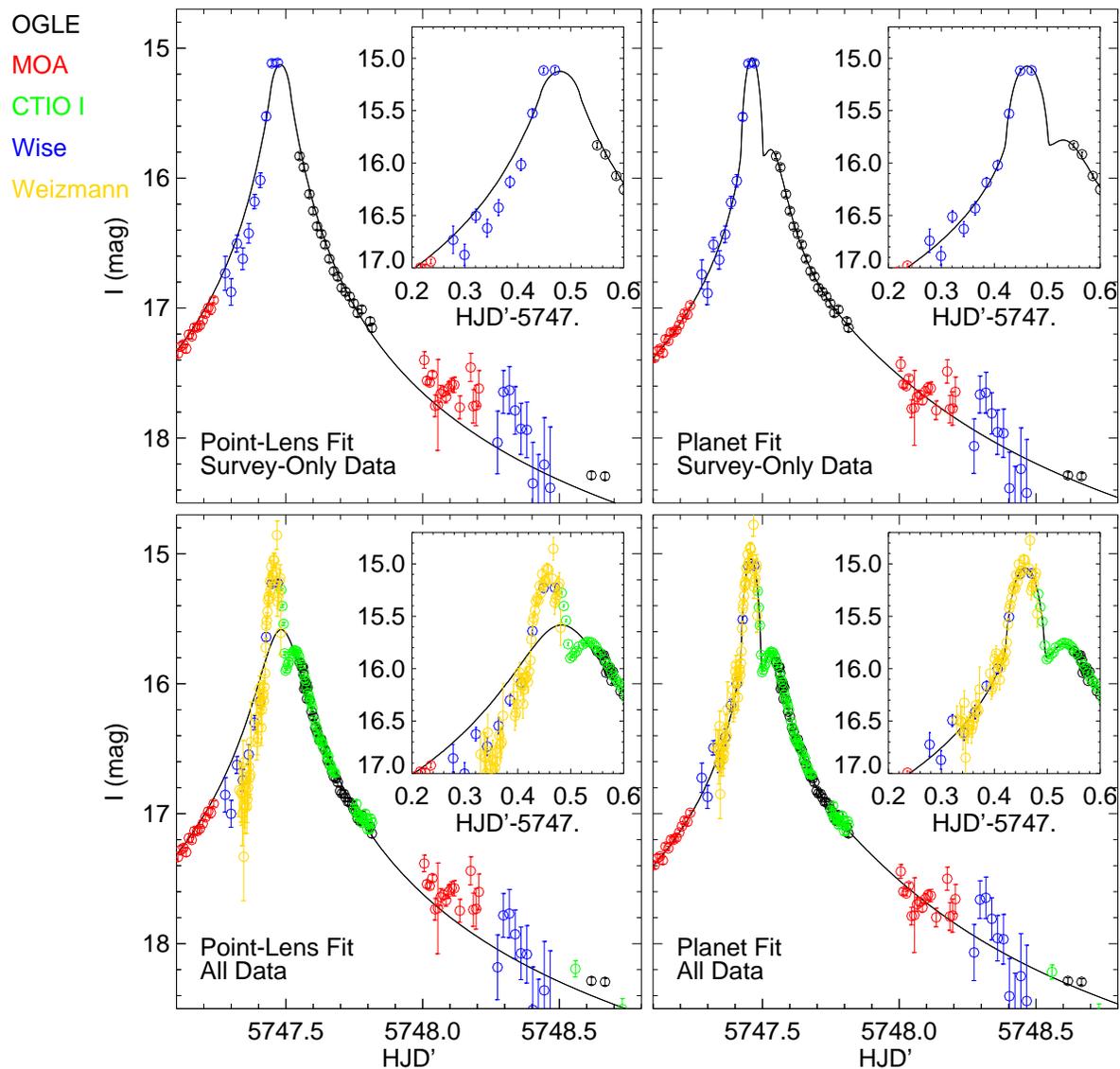}
\caption{Comparison of point-lens fits ({\it left}) and planet fits
({\it right}) for ``survey-only'' data, ({\it top}) and all data ({\it
bottom}). In both cases, the planet fit is clearly better than the
point lens fit, but the difference is more significant when followup
data are included. Note that for ``survey-only'' data the OGLE data
have been thinned out to reflect the typical survey cadence. \label{fig:lccomp}}
\end{figure}

We repeat the analysis on this survey-only data set beginning with a
broad grid search and then refining the solution following the same
procedure used for analyzing the complete data set. We find that even
without flux-alignment constraints, the global search isolates
solutions in the general neighborhood of the solution found from the
full data set. The fits to the survey-only data set are compared to
fits with all data in Figure \ref{fig:lccomp}.  Here, the
$\Delta\chi^2$ of the fit compared to a point lens fit for the
survey-only data is 487, nearly all of which comes from data in the time-span
shown in Figure \ref{fig:lccomp} $5747.1<t(HJD^{\prime})<5748.8$. This
is smaller than the $\Delta\chi^2$ of any published microlensing
planet. However, the parameters of the fit are well constrained with
errors only a factor of 1.5-2 larger compared to fits with the full
data set. Applying the flux-alignment constraint to this model
confirms its validity, i.e., it does not appreciably change the
solution (see Table~\ref{tab:models}). In this case, it is clear that
the survey data are sufficient to robustly detect and characterize the
planet.

In order to push farther into the limits of detectability, we also
analyze this event without the Wise data, since those data contain
most of the deviation from the underlying point lens. With only the
MOA and thinned OGLE data, we find $\Delta\chi^2\sim70$ between the
2-body lens and point lens models. We note that the point lens model
has an unreasonably large value of parallax, $\pi_{\rm E}\sim 20$,
making it somewhat suspicious. Without parallax, the $\Delta\chi^2$
between the point lens and 2-body fit increases to
$\Delta\chi^2\sim170$. Although this is a factor of 3 smaller than the
$\Delta\chi^2$ with the Wise data, the constraints on the planet-star
mass ratio are still broadly confined to be planetary, assuming
$M_L\lesssim0.5 $M$_{\odot}$, with a $3\sigma$ range of $0.001\lesssim
q\lesssim 0.025$.  However, because of the small $\Delta\chi^2$, it
appears unlikely to us that a planetary detection would have been
claimed from solely the MOA and OGLE data even though the solution is
formally well-constrained.

{\section{Physical Properties of the Event}
\label{sec:eventprop}}

Since finite source effects are measured in this event, we can
determine the angular size of the Einstein ring, $\theta_{\rm E}$, and
the lens-source relative proper motion, $\mu$. First, we
estimate the angular size of the source, $\theta_{\star}$, from the
observed color and magnitude. We transform the $(V-I)_{S,0}$ color to $(V-K)$
using the dwarf relation from \citet{Bessell88}. Then we use the
$(V-K)$ surface brightness relations from \citet{Kervella04} to find
$\theta_{\star}=0.42\pm 0.03\,\mu$as. From this we derive the
lens-source relative proper motion and angular Einstein radius,
\begin{equation}
\mu = \frac{\theta_{\star}}{t_{\star}}= 4.3\pm 0.3\,
{\rm mas\,yr^{-1}};
\quad
\theta_{\rm E} = \mu t_{\rm E} = 0.26\pm 0.02\,{\rm mas}.
\label{eqn:muthetae}
\end{equation}
The uncertainties in these quantities come from a variety of
factors. Specifically, the uncertainties in the Galactocentric
distance, $R_0$, and the measured intrinsic brightness of the red
clump, the centroiding of the red clump from the CMD, and uncertainty
in the surface brightness relations. The uncertainty contributed by
the surface brightness relations is 0.02 mag, and the uncertainties
from the other factors are given in Section \ref{sec:cmd}. The
contribution of these factors can be understood from their
relationship to $\theta_{\star}$ \citep{Yee09}:
\begin{equation}
\theta_{\star}=\frac{\sqrt{f_S}}{Z},
\end{equation}
where $f_S$ is the source flux from the microlensing model and $Z$
captures all other factors. Taking account of all factors mentioned
above, we find $\sigma(Z)/Z= 8\%$.  Since the statistical error in
$f_S^{1/2}$ is only 2.3\%, the error in $Z$ completely dominates the
uncertainty in $\theta_*$.  In general, the error in $f_S$ propagates
in opposite directions for $\theta_{\rm E}$ and $\mu$ \citep{Yee09}.
However, in the present case, since this error is small, the
fractional error in these quantities is simply that of $Z$, as
indicated in Equation~(\ref{eqn:muthetae}).

{\section{Properties of the Lens}
\label{sec:physical}}

\subsection{Limits on the Lens Brightness}
We can use the observed brightness of the event to place constraints
on the lens mass. Since the source and lens are superposed, any light
from the lens should be accounted for by the blend flux, $f_{B,i}$,
which sets an upper limit on the light from the lens. The unmagnified
source is not seen in the OGLE data. From examination of an OGLE image
at baseline with good seeing, we estimate the upper limit of the blend
flux to be $I_{B,0} \geq 17.77$ based on the diffuse background light
and assuming that the reddening is the same as the red clump. Assuming
all of this light is due to the lens, the absolute magnitude of the
lens is
\begin{equation}
M_{I,L} > I_{B,0} + (A_{I,S} -A_{I,L}) - 5\log {D_L\over 10\,{\rm pc}}
= 3.25 + (A_{I,S} - A_{I,L}) + 5\log {R_0\over D_L},
\end{equation}
where $A_{I,S}$ and $A_{I,L}$ are the reddening toward the source and
lens, respectively, and $D_L$ is the distance to the lens.  Since the
lens must be in front of the source, we have $A_{I,S}\geq A_{I,L}$.
Moreover, the lens should be closer than $R_0$ (or at any rate, not
much farther). Hence, $M_{I,L}\geq3.25$ is a conservative lower
limit. From the empirical isochrones of \citet{An07}, this absolute
magnitude corresponds to an upper limit in the lens mass of $M_L\leq
1.2 M_{\odot}$. We conclude from these flux-alignment constraints that either
the lens is a main sequence star or, if it is more massive than our
upper limit of $1.2 M_{\odot}$, then it must be a stellar remnant such
as a very massive white dwarf or a neutron star.

We can use our measurement of $\theta_{\rm E}$ to estimate the distance to
the lens based on its mass:
\begin{equation}
D_L=\left(\frac{\theta_{\rm E}^2}{\kappa M_L}\frac{1}{\rm AU}+\frac{1}{D_S}\right)^{-1} \mathrm{\,with\,} \kappa\equiv\frac{4G}{c^2 \mathrm{AU}}=8.14\mathrm{\,mas\,} M_{\odot}^{-1},
\end{equation}
where $D_S$ is the distance to the source. If we assume the source is
at 8 kpc (i.e., about 0.1 mag behind the mean distance to the clump at
this location) and $M_L=1.2 M_{\odot}$, we find $D_L=7.6\,$kpc. Hence,
the lens could be an F/G dwarf or stellar remnant in the Bulge, or it
could be a late-type star closer to the Sun.

\subsection{Bayesian Analysis}
\label{sec:bayes}

Similar to \citet{Alcock97} and \citet{Dominik06}, we estimate the
mass of the lens star and its distance using Bayesian analysis
accounting for the measured microlensing parameters, the brightness
constraints on the lens, and a model for the Galaxy. The mathematics
are similar to what is described in Section 5 of \citet{Batista11},
although the implementation is fundamentally different because we do
not have meaningful parallax information. Specifically, we perform a
numerical integral instead of applying the Bayesian analysis to the
results of the MCMC procedure. We begin with the rate equation for
lensing events:
\begin{equation}
\frac{d^4\Gamma}{dD_LdM_Ld^2\mu} = \nu(x,y,z)(2R_{\rm E})v_{\rm
rel}f({\bf\mu})g(M_L),
\end{equation}
where $\nu(x,y,z)$ is the density of lenses, $R_{\rm E}$ is the
physical Einstein radius, $v_{\rm rel}$ is the lens-source relative
velocity, $f({\bf\mu})$ is the weighting for the lens-source relative
proper motion, and $g(M_L)$ is the mass function. The vector form of the lens-source
relative proper motion is ${\bdv \mu}$, which can be described by a
magnitude, $\mu$, and an angle, $\phi$, such that $d^2\mu=\mu d\mu
d\phi$. We transform variables (see \citealt{Batista11}) to find
\begin{equation}
\frac{d^4\Gamma}{dD_Ld\theta_{\rm E}dt_{\rm E}d\phi} =
\frac{2D_L^2\mu^4\theta_{\rm E}}{\kappa\pi_{\rm rel}}
\nu(x,y,z)f({\bdv\mu})g(M_L).
\end{equation}
To find the probability density functions for the lens, we
integrate this equation over the variables $\theta_E$ and $\phi$,
using a Gaussian prior for $\theta_E$ with the values given in
Eq. (\ref{eqn:muthetae}) and a flat prior for $\phi$. We calculate
$\mu$ from $t_{\rm E}$ and $\theta_{\rm E}$ using Equation
(\ref{eqn:muthetae}). We also integrate over $D_S$, which appears
implicitly in $\pi_{\rm rel}$ and $f(\mu)$. For $D_S$, we include a
prior for the density of sources based on our Galactic model (see
below) assuming the source is in the Bulge.

Three functions remain to be defined\footnote{We will neglect
constants of proportionality as they are not relevant to a likelihood
analysis.}: $\nu(x,y,z)$, $f({\bdv\mu})$, and $g(M_L)$. As in
\citet{Batista11}, we assume $g(M)\propto M^{-1}$. For the proper
motion term, we follow Equation~(19) of \citet{Batista11}:
\begin{equation}
f_{\mu} \propto \frac{1}{\sigma_{\mu,
{N_{\rm gal}}}\sigma_{\mu, E_{\rm gal}}}\exp{\left[-\frac{(\mu_{N_{\rm gal}}-\mu_{{\rm exp},N_{\rm gal}})^2}{2\sigma_{\mu,
N_{\rm gal}}^2}-\frac{(\mu_{\rm E_{\rm gal}}-\mu_{{\rm exp},E_{\rm gal}})^2}{2\sigma_{\mu, E_{\rm gal}}^2}\right]}.
\end{equation}
Note that the variables in $f_{\mu}$ are given in Galactic coordinates
rather than Equatorial coordinates. The transformation between the two
is simply a rotation by 60$^{\circ}$. Still working in Galactic
coordinates, the expected proper motion, ${\bdv\mu}_{\rm exp}$, takes
into account the typical motion of a star in the Disk, ${\bf v}$, and
the motion of the Earth during the event, ${\bf
v}_{\oplus}=(v_{\oplus, N_{\rm gal}},v_{\oplus, E_{\rm
gal}})=(-0.80, 28.52)\,\mathrm{km\, s^{-1}}$,
\begin{equation}
{\bdv \mu}_{\rm exp} = \frac{{\bf v}_{L}-({\bf v}_\odot+{\bf v}_\oplus)}{D_L}
- \frac{{\bf v}_{S}-({\bf v}_\odot+{\bf v}_\oplus)}{D_S},
\end{equation}
where ${\bf v}_\odot = (7,12)\,{\rm km}\,{\rm s}^{-1} + (0,v_{\rm
rot})$ and $v_{\rm rot}=230\,{\rm km}\,{\rm s}^{-1}$.  For the Disk we
use ${\bf v}=(0,v_{\rm rot}-10\,\mathrm{km\, s^{-1}})$ and
${\bdv\sigma}=(\sigma_{\mu, N_{\rm gal}},\sigma_{\mu, E_{\rm
gal}})=(20,30)\,\mathrm{km\, s^{-1}}$, and for the Bulge ${\bf
v}=(0,0)\,\mathrm{km\, s^{-1}}$ and ${\bdv\sigma}=(\sigma_{\mu, N_{\rm
gal}},\sigma_{\mu, E_{\rm gal}})=(100,100)\,\mathrm{km\, s^{-1}}$. 

For the stellar density $\nu(x,y,z)$, we use the model from
\citet{Han03} including a bar in the Bulge. We assume the Disk has
cylindrical symmetry with a hole of radius 1 kpc centered at $R_0=8\,$
kpc. We limit the Bulge to $5<D<10\,$ kpc, where $D$ is the distance
from the observer along the line of sight.

For the Bayesian analysis, we use $t_E = 21.7$ days measured from the
microlensing fit to the light curve. We also have the constraint from
the lens brightness that $M_L=\theta_E^2/(\kappa\pi_{\rm rel})<1.2
M_{\odot}$. This analysis implicitly assumes that the lens is a main
sequence star. The lens could be a stellar remnant, although this is
much less likely because of their smaller relative space density. The
possibility that the lens is a stellar remnant could be tested several
years from now when the source and lens have moved sufficiently far
apart so as to be separately resolved, i.e., in roughly
$10(\lambda/1.6 {\rm \mu m})(D_{\rm tel}/10 {\rm m})^{-1}$ years,
where $\lambda$ is the wavelength of the observations and $D_{\rm
tel}$ is the diameter of the telescope used, assuming the observations
are diffraction limited (for a discussion of detecting light from a
main sequence lens, see the Appendix).

\begin{figure}
\includegraphics[height=0.7\textheight]{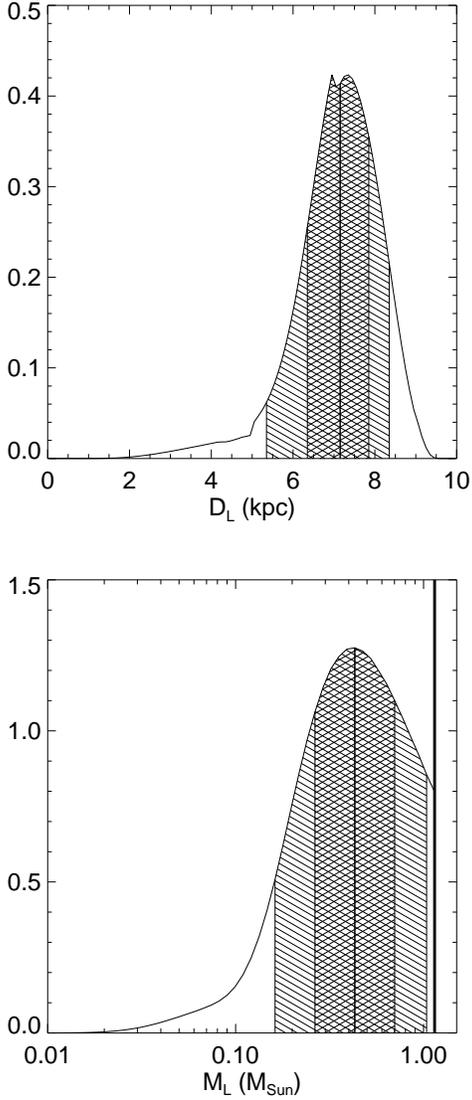}
\caption{Probability densities for the lens (host) as function of its
distance ({\it top}) and mass ({\it bottom}). The vertical scale is
set so that the total area under each curve is equal to one. Masses
$M_L>1.2\,M_\odot$ are excluded by the flux-alignment constraint on
the lens brightness (bold vertical line). The 68\% and 90\% confidence
intervals about the median are indicated by the shaded regions. The
discontinuities in the slope of probability distribution for the lens
distance arise from overlap between the Disk and Bulge stellar density
distributions. From the Galactic model priors, there is a significant
probability that the host is an M-dwarf. High-resolution imaging could
confirm or contradict this by direct detection of the lens light (see
Appendix).
\label{fig:bayes}}
\end{figure}

The results of the Bayesian analysis are shown in Figure
\ref{fig:bayes}. We find that if the lens is a main sequence star, its
mass is $M_L = 0.43^{+0.27}_{-0.17}\,M_\odot$ and its distance is
$D_L=7.15\pm 0.75\,$kpc (median and 68\% confidence interval). Hence
the planet mass is $m_p=2.4^{+1.5}_{-0.9}\,M_{\rm Jup}$.  In the close
solution, the projected separation is sharply peaked at $r_{\perp}=s
D_L\theta_{\rm E}=\,1.0\pm 0.1\,$AU.  However, the wide solution,
which is not strongly disfavored, gives an alternative
$r_{\perp}=3.4\pm 0.4\,$AU. If we assume $a\sim r_{\perp}$, the planet
would have a period of $\sim1.5$ or $\sim8$ years.

{\section{Discussion}
\label{sec:discuss}}

\subsection{Implications for Planet Formation Theory}

The lens in MOA-2011-BLG-293 consists of a super-Jupiter orbiting a
probable M dwarf. The projected separation of the planet from the star
is at most a few AU, making it difficult to form {\it in situ} if the
host is indeed an M dwarf. Core accretion theory makes a general
prediction that massive Jovian planets around M dwarfs should be rare
\citep{Laughlin04,Ida05}. While gravitational instability can form
large planets around M dwarfs \citep{Boss06}, these typically form
farther out, so if the planet formed by this mechanism, it would
either be required to have migrated significantly or the projection
effects must be severe. In the Appendix, we discuss how Adaptive
Optics (AO) observations can confirm the microlensing measurement of
the host mass or at least place upper limits on the host mass that
would confine it to the M dwarf regime. Additionally, it should be
noted that this planet joins a growing number of massive planets
orbiting stars likely to be M dwarfs discovered by microlensing (see
Appendix) and radial velocity (see \citealt{Bonfils11} for a summary
and also \citealt{Johnson11}).

{\subsection{The Ongoing Importance of Followup}} 

Even with high-cadence surveys, followup data remain important for the
interpretation of individual events. In this case, the event was high
magnification with a faint source, undetected at baseline. This meant
that the time period when the event was observable, to surveys and
followup, was very brief. We have shown that followup data can vastly
increase the signal and provide redundancy in the light curve
coverage, which protects against weather. In Section
\ref{sec:surveyonly}, we demonstrated that with the survey data, even
though the break in the light curve is missed because of bad weather,
the same solution is recovered for MOA-2011-BLG-293, albeit at much
lower significance than when all the data are included
($\Delta\chi^2\sim500$ compared to $\Delta\chi^2\sim5400$). This is
somewhat surprising since it is conceivable that because of the
degeneracies possible for central caustic type events, the loss of
this feature would leave the models relatively unconstrained or allow
alternative solutions to the light curve. The followup data are
beneficial in this case because in addition to adding to the
signal-to-noise, they trace the sharp feature seen in the model light
curves, increasing our confidence that the correct model has been
found.

It is likely that without the real-time discovery of this event in the
MOA data and the subsequent high magnification alert from $\mu$FUN,
the planet would have remained undiscovered as of this writing. The
planetary anomaly only becomes detectable when data from all three
survey telescopes are combined, which requires systematically
reducing, combining, and searching all of the survey data, preferably
using a difference imaging reduction in order to detect events with
faint sources like this one. Routine reduction of all survey data is
planned for the current OGLE/MOA/Wise survey, but has not yet been
fully implemented.

The multi-band data taken by followup groups can also be important for
interpreting microlensing events. In order to determine the physical
characteristics of the lens from the microlens variables, we used the
$(V-I)$ color of the source to estimate its angular size, thus
providing a physical scale for the lens system. We also used the
$(V-I)$ color to inform our choice of limb-darkening coefficients for
our model. Additionally, $H$-band data are important for comparison to
AO observations that may be used to improve constraints on the lens
mass as discussed for this event in the Appendix. $H$-band data are
routinely taken as part of followup observations at CTIO but not part
of the planned surveys.  The OGLE survey regularly takes $V$-band data
every few days when the weather is good, and the Wise survey is
planning similar observations for future seasons. Provided the event
timescale is long, this will result in several points taken when the
source is substantially magnified.

For the present case, the highly magnified part of the event is
brief, and OGLE only obtained one $V$ band point over peak, which was
taken deliberately as part of followup observations. Because the
$V$-band data were only taken as part of followup, we need to consider
the effect of excluding these data in the context of a pure survey
detection. In principle, the $(V-I)$ color could be estimated
following the method in \citet{Gould10a}, which takes advantage of the
difference in the OGLE and MOA bandpasses to estimate the $(V-I)$
color from $R_{\rm MOA}-I_{\rm OGLE}$. For this event, the uncertainty
of this measurement from the fits to survey-only data is
$\sigma_{R-I}=0.01$ leading to a $(V-I)$ uncertainty of
$\sigma_{V-I}=0.04$. In this case, the precision is not much worse
than what we found from the standard technique using CTIO $V$- and
$I$-band photometry, so the lack of survey $V$-band data would not
have a major effect.

Finally, although this event clearly shows that a planet is detected
at $\Delta\chi^2\sim500$ without followup data, this is smaller than
the $\Delta\chi^2$ of any published microlensing planet, underlining
the fact that low-significance signals have not been systematically
explored. Events like MOA-2011-BLG-293 that are robustly characterized
with followup data but with weaker signals in the survey data can be
used to probe lower $\Delta\chi^2$ signals and inform our
understanding of the limits of what is detectable. For example, if we
analyze a large number of events with a range of signal strengths in
the survey data, we could determine a $\Delta\chi^2$ threshold for
which a known signal, seen in the complete data set including
followup, can no longer be distinguised from the noise. Additionally,
the results for central caustic events will help us understand a
$\Delta\chi^2$ threshold below which the model degeneracies mean that
the ``correct'' solution, as determined from the complete data set, can
no longer be recovered. We might also require that these
events can be well-characterized, i.e., that degenerate central
caustic models can be sufficiently disentangled so that the mass ratio
is well-constrained\footnote{We will leave the exact definition of
this phrase to future investigations but suggest that it might be
along the lines of constraining the mass ratio to an order of
magnitude at $2\sigma$.}. Understanding these thresholds will be
important for analyzing large samples of events to study the planets
as a population rather than individual discoveries. By analyzing a
large sample of events similar to MOA-2011-BLG-293, we can empirically
determine appropriate $\Delta\chi^2$ thresholds or investigate other
statistics for both detecting planets in all microlensing events and
characterizing them in central caustic events.

\acknowledgements{The OGLE project has received funding from the
European Research Council under the European Community's Seventh
Framework Programme (FP7/2007-2013) / ERC grant agreement
no. 246678. The MOA collaboration acknowledges the support of grants
JSPS22403003 and JSPS23340064. Work by J.C. Yee is supported by a
National Science Foundation Graduate Research Fellowship under Grant
No. 2009068160. A. Gould and J.C. Yee acknowledge support from NSF
AST-1103471. B.S. Gaudi, A. Gould, and R.W. Pogge acknowledge support
from NASA grant NNG04GL51G.  A. Gal-Yam acknowledges support by the
Benoziyo Center for Astrophysics, and by the Lord Sieff of Brimpton
Memorial Fund. Work by C. Han was supported by Creative Research
Initiative Program (2009-0081561) of National Research Foundation of
Korea. T. Sumi acknowledges support from JSPS23340044. Work by
S. Dong was performed under contract with the California Institute of
Technology (Caltech) funded by NASA through the Sagan Fellowship
Program.}

\appendix
{\section{Appendix: Possible Constraints from AO Observations}
\label{sec:discuss1}}

Of the 13 previously published microlensing planets, two are very
likely to be super-Jupiters orbiting M dwarfs\footnote{Two more could be
marginally included in this category \citep{Bennett06,Dong09}.}.  In
both cases, high-resolution imaging from space or the ground
was needed to complete these determinations. OGLE-2005-BLG-071 has a
mass ratio, $q=7.4\pm 0.4\times 10^{-3}$ \citep{Udalski05,Dong09_071},
similar to that of MOA-2011-BLG-293 analyzed here.  \citet{Dong09_071}
subsequently combined {\it Hubble Space Telescope (HST)} and light
curve data to determine the host mass $M_L = 0.46\pm 0.04\, M_\odot$,
implying that this was an $m_p\sim 3.5\,M_{\rm Jup}$ planet orbiting
an M dwarf. \citet{Batista11} found an even higher mass ratio for
MOA-2009-BLG-387 of $q=0.0132\pm 0.003$.  Their marginal detection or
upper limit on the lens flux from 8m-class adaptive optics (AO)
observations allowed them to place an upper limit $M<0.5\,M_\odot$
(90\% confidence) on the host, with a median estimate of $M_L =
0.19\,M_\odot$, and so $m_p = 2.6\,M_{\rm Jup}$.

Based on our Bayesian analysis (Section \ref{sec:bayes}), it is likely
that MOA-2011-BLG-293L is another case of a super-Jupiter orbiting an
M dwarf. As we now show, 8m-class adaptive optics (AO) observations
could clarify the nature of the system. The main uncertainty in an AO
measurement of the lens flux is the uncertainty in the source flux,
since the two objects are superposed unless many years have passed
since the event. Because an alternate model for the baseline flux
exists in which the flux is not constant (see Section \ref{sec:tefs}),
we cannot be certain it is possible to reliably measure the blended
flux. Nevertheless, for the sake of argument, we are going to assume
that the baseline flux is stable, so the uncertainty in $f_S$ is
dominated by the statistical errors from the MCMC procedure. This
assumption can be tested after OGLE-IV has collected several more
years of baseline data on the event. If these data confirm that the
baseline is stable, the calculations in this section are
applicable. If not, the limits estimated here are overly optimistic.

As mentioned in Section~\ref{sec:eventprop}, the statistical error in $f_S$
is 4.6\%, which is due primarily to correlations of the source flux
with other model parameters, rather than errors from fitting the
light curve to an individual model.  Thus, the $H$-band flux (in units
of the instrumental scale of CTIO $H$-band images) is known to
essentially the same precision.  Using standard techniques
\citep{Janczak10,Batista11}, this flux scale can be aligned to the AO
flux scale to a few percent.  Hence, the AO source flux $f_{H,S}$ can be
known to about 7\%.  This means that excess light due to the lens can
be securely detected at the $3\,\sigma$ level, provided that $\Delta
H\equiv H_L - H_S < 1.7$ mag.  This quantity can be related to the
physical properties of the lens and source by
\begin{equation}
\Delta H \equiv H_L - H_S = \Delta M_H + \Delta A_H + \Delta {\rm Dmod}
\label{eqn:deltah}
\end{equation}
where $\Delta M_H\equiv M_{H,L}-M_{H,S}$,  
$\Delta A_H\equiv A_{H,L}-A_{H,S}$,  and
$\Delta {\rm Dmod} \equiv 5\log(D_L/D_S)$.  Now, in the regime we will
be considering, it is very likely that $\Delta A_H=0$, but in any
case $\Delta A_H\leq 0$, since the lens is in front of the source.
Hence, we can conservatively ignore this term.  The last term is
\begin{equation}
\Delta {\rm Dmod} = -5\log\biggl(1 + {D_S\over {\rm AU}}
{\theta_{\rm E}^2\over \kappa M_L} \biggr)
\rightarrow -0.3{D_S\over 8\,{\rm kpc}}\,{0.45\,M_\odot\over M_L}
\label{eqn:deltadmod}
\end{equation}
where in the second step we have inserted
Equation~(\ref{eqn:muthetae}) and kept only the first term of the
Taylor expansion of the logarithm\footnote{This approximation assumes that
$(D_S/D_L - 1)\ll 1$, where $D_L$ is implicitly embedded in
$\theta_{\rm E}$.}.  Hence, the lens will be detectable provided
\begin{equation}
M_{H,L} \la M_{H,S} + 2.0,
\label{eqn:lensdetect}
\end{equation}
where $2.0 = (\Delta H-\Delta {\rm Dmod})$ assuming $D_S=8\,$kpc and
 $M_L=0.45\,M_\odot$.  From its color and magnitude, the source is a
 late G/early K Bulge dwarf, so $M_{H,S}\sim 4.2$.  Hence, all dwarfs
 $M_{H,L}\la 6.2$ are detectable, which corresponds to $M_L\ga
 0.43\,M_\odot$.

Such a detection or upper limit would not be absolutely secure.
Detected light could in principle be due to a companion to the
lens or source, or an unrelated star in this crowded, low-latitude
Bulge field.  Additionally, an upper limit on the lens flux could be attributed to
a remnant host rather than a late M dwarf.  Nevertheless, the
probabilities of these alternative interpretations can be quantified,
and it is important to do so in order to estimate the frequency of
very massive planets orbiting M dwarfs.

\bibliographystyle{/home/morgan/jyee/tex/apj}



\begin{deluxetable}{lllcr}
\tablecaption{Data\label{tab:data}}
\tablehead{\colhead{}&\colhead{}&\multicolumn{2}{c}{Error Renormalization Coefficients}&\colhead{}\\
\colhead{Observatory}&\colhead{Filter}&\colhead{$k$}&\colhead{$e_{\rm min}$}&\colhead{$N_{\rm data}$}\\
}
\startdata
OGLE     & I          &1.75 & 0.01 &274\tablenotemark{a}\\
MOA      & MOA-Red    &1.25 & 0.0  &78\tablenotemark{b} \\
CTIO     & I          &1.56 & 0.0  &63  \\
Wise     & I          &1.57 & 0.0  &49  \\
Weizmann & I          &1.74 & 0.0  &54  \\
CTIO\tablenotemark{c} & V &\nodata&\nodata&9\\
\enddata
\tablenotetext{a}{$N_{\rm data}$ after binning.}
\tablenotetext{b}{$N_{\rm data}$ after binning. Restricted to
$5743.5<t(HJD^{\prime})<5749.5$.}
\tablenotetext{c}{These data were not used in light curve
modeling. They were only used to determine the color
of the source.}
\tablecomments{The properties of each data set are given along with the
error renormalization coefficients used to rescale the error bars (see
Sec. \ref{sec:error}).}
\end{deluxetable}

\begin{deluxetable}{lllllllllllll}
\tablecaption{Lightcurve ``Invariants''\label{tab:invariants}}
\tablehead{\colhead{Model}&\colhead{$t_{\rm eff}$}&\colhead{$f_{\rm
lim}$}&\colhead{$t_{\star}$}&\colhead{$qt_{\rm E}$}\\
\colhead{}&\colhead{(days)}&\colhead{(\tablenotemark{a})}&\colhead{(days)}&\colhead{(days)}
}
\startdata 
           close& 0.0756(5) &   9.81(6) & 0.0355(3) &  0.115(2) &\\
\hline
      close with& 0.0754(5) &   9.82(7) & 0.0355(3) &  0.115(2) &\\
flux constraints&\\
\hline
      close with& 0.0748(7) &  9.95(11) & 0.0355(7) &  0.113(2) &\\
        parallax&\\
\hline
            wide& 0.0754(5) &   9.85(7) & 0.0355(3) &  0.116(2) &\\
\hline
       wide with& 0.0754(5) &   9.85(7) & 0.0355(3) &  0.116(2) &\\
flux constraints&\\
\hline
\hline
     survey only&  0.075(2) &   10.1(3) &  0.039(3) &  0.109(7) &\\
\hline
     survey with&  0.076(2) &    9.9(3) &  0.040(3) &  0.110(8) &\\
flux constraints&\\
\hline
\enddata
\tablenotetext{a}{$f_{\rm lim}\equiv f_{S, {\rm OGLE}}/u_0$,
where $f_{S, {\rm OGLE}}=1$ corresponds to a magnitude $I=18$, so
$f_{\rm lim}$ has units of flux in this system.}
\tablecomments{Comparing the invariants of the lightcurve
(Sec. \ref{sec:invariants}) shows that they are robustly measured
both in terms of their uncertainties and their variation among
models.} 
\end{deluxetable}

\begin{deluxetable}{lllllllllllll}
\tablecaption{Model Parameters\label{tab:models}}
\rotate
\tabletypesize{\scriptsize}
\setlength{\tabcolsep}{0.04in}
\tablehead{\colhead{Model}&\colhead{$\chi^2$}&\colhead{$t_0-5747.$}&\colhead{$u_0$}&\colhead{$t_{\rm E}$}&\colhead{$\rho$}&\colhead{$\alpha$}&\colhead{$s$}&\colhead{$q$}&\colhead{$\pi_{E,N}$}&\colhead{$\pi_{E,E}$}&\colhead{$\frac{f_{S,Wise}}{f_{S,CTIO}}$}&\colhead{$\frac{f_{S,Weizmann}}{f_{S,CTIO}}$}\\
\colhead{}&\colhead{}&\colhead{(HJD$^{\prime}$)}&\colhead{}&\colhead{(days)}&\colhead{}&\colhead{($^{\circ}$)}&\colhead{}&\colhead{}&\colhead{}&\colhead{}&\colhead{}&\colhead{}}
\startdata
           close&  658.9377& 0.4935(7) & 0.0035(2) & 21.67(96) &0.00164(7) &  221.3(5) &  0.548(6) & 0.0053(2) &     0.(.) &     0.(.) &  0.979(9) &   1.09(2) \\
\hline
      close with&  662.0860& 0.4935(6) & 0.0035(2) & 21.75(95) &0.00163(7) &  221.3(5) &  0.548(5) & 0.0053(2) &     0.(.) &     0.(.) &  0.990(4) &   1.08(1) \\
flux constraints\\
\hline
      close with&  655.5644& 0.4924(9) & 0.0035(2) & 21.24(95) &0.00168(8) &  221.5(6) &  0.552(6) & 0.0054(2) &  1.7(1.1) & -2.4(1.5) &   0.94(2) &   1.04(3) \\
        parallax\\
\hline
            wide&  662.8497& 0.4931(7) & 0.0034(2) & 22.49(98) &0.00158(7) &  221.1(5) &   1.83(2) & 0.0052(2) &     0.(.) &     0.(.) &   0.98(1) &   1.08(2) \\
\hline
       wide with&  665.9169& 0.4931(6) & 0.0033(1) & 22.64(98) &0.00157(7) &  221.1(5) &   1.83(2) & 0.0051(2) &     0.(.) &     0.(.) &  0.988(5) &   1.07(1) \\
flux constraints\\
\hline
\hline
     survey only&  497.3160&  0.492(1) & 0.0038(2) & 19.8(1.0) & 0.0020(2) &    218(1) &   0.55(1) & 0.0055(4) &     0.(.) &     0.(.) &                        \\
\hline
survey only with&  498.8901&  0.493(1) & 0.0038(2) & 20.0(1.0) & 0.0020(2) &    218(1) &   0.55(2) & 0.0055(4) &     0.(.) &     0.(.) &                        \\
flux constraints\\
\hline
\enddata
\tablecomments{The mean and root mean square errors for the parameters of
each model are given along with the $\chi^2$ for that model. The fits
with ``survey only'' use only a subset of data representative of what
would have been obtained without additional followup. Note that the
parameters of these fits are very similar to the parameters of the
other fits, but with slight increases in their uncertainties.}
\end{deluxetable}

\end{document}